\makeatletter\let\ifGm@compatii\relax\makeatother
\documentclass[pre,aps,twocolumn,floatfix,superscriptaddress]{revtex4}
\usepackage{eso-pic,calc}
\usepackage{graphicx}
\usepackage{amsmath, amsthm, amssymb, mathrsfs}
\usepackage{epsfig}
\usepackage{latexsym}
\usepackage{bm}

\def\beqr{\begin{eqnarray}}
\def\eqnr{\end{eqnarray}}
\def\beq{\begin{equation}}
\def\bc{\begin{center}}
\def\ec{\end{center}}
\def\eqn{\end{equation}}

\def\rmp#1#2#3{{ Rev. Mod. Phys.} {\bf #1}, #2 (#3)}
\def\prl#1#2#3{{ Phys. Rev. Lett.} {\bf #1}, #2 (#3)}

\def\pre#1#2#3{Phys. Rev. E {\bf #1}, #2 (#3)}

\def\pnas#1#2#3{Proc. Natl. Acad. Sci. (USA) {\bf #1}, #2 (#3)}

\def\natphys#1#2#3{Nat. Phys. {\bf #1}, #2 (#3)}
\def\sc#1#2#3{Science {\bf #1}, #2 (#3)}

\begin{document}
\title{Scaling in simple continued fraction}
\author{Avinash Chand Yadav}
\affiliation{Department of Physics and Astronomical Sciences, Central University of Jammu, Samba 181 143, India}

\begin{abstract}
{We consider a class of real numbers, a subset of irrational numbers and certain mathematical constants, for which the elements in the simple continued fraction appears to be random. As an illustrative example, one can consider $\pi = \{x_0, x_1, x_2, \dots x_n\}$, where $x$'s are the continued fraction elements computed with an exact value of $\pi$ up to $N$ precision. We numerically compute probability distribution for the elements and observe a striking power-law behavior $P(x)\sim x^{-2}$. The statistical analysis indicates that the elements are uncorrelated and the scaling is robust with respect to the precision.  Our arguments reveal that the underlying mechanism generating such a scaling may be sample space reducing process. }
\end{abstract}

\maketitle
\section{Introduction} 
In many systems, the observable of interest is typically a random variable satisfying probability distribution of the power-law form
\begin{equation}
P(x)\sim x^{-\tau},
\label{eq1}
\end{equation}
where $\tau$ is a critical exponent \cite{Newman_2005, Sornette_2006, Miguel_2018}. Striking examples have been seen in a diverse settings, ranging from city size  distribution to frequency distribution of words in a typical text, and critical avalanches observed in sandpiles \cite{Bak_1996, Bak_1987, Yadav_2012} or neuronal  activity \cite{Levina_2007, Manchanda_2013, Jonathan_2017}. The Eq.~(\ref{eq1}) remains unchanged under the scaling of its argument. The scale invariance implies  lack of a characteristic length scale. In other words, the events of all size are possible; while the smaller size events appear more frequent,  the larger size events occur rarely, but may be catastrophic.

Uncovering a peculiar system exhibiting emergent scaling  as well as origin of the power-law remains a topic of continuing interest in the study of complex systems. Note that the Gaussian statistics emerge naturally as a consequence of a distinct origin that is `central limit theorem'. On the contrary, the occurrence of power-law may have a few different dynamical routes. The most common underlying mechanisms include: Criticality \cite{Djordje_2011, Knecht_2012, Munn_2018, Kinouchi_2006}, self-organized criticality \cite{Bak_1996, Bak_1987, Yadav_2012}, first passage properties for random walk \cite{Redner_2001, Dickman_2001, Avinash_2018}, preferential process \cite{Albert_1999},  multiplicative stochastic process with constraint \cite{Levy_1996, Takayasu_1997, Yamamoto_2012, Yamamoto_2014}, and sample space reducing (SSR) process \cite{Murtra_2015, Murtra_2017, Murtra_2018, Yadav_2016, Yadav_2017}.

For certain processes such as sentence formation to fragmentation or technological innovation to biological extinction, the sample space, the set of accessible states, dynamically evolves as a function of time. In particular, the sample space reducing stochastic process is of considerable interest, as this offers an explanation of Zipf's law. The Eq.~(\ref{eq1}) describes the standard Zipf's law, if $x$ is a positive integer and $\tau = 2$. However, a proper balance between SSR (relaxing) and expanding (driving) processes can explain other interesting distributions \cite{Murtra_2018}.

A power-law probability function with the critical exponent 2 has been observed in a variety of contexts with different dynamical origin. It is worthwhile to mention few examples to see a contrast. Instances include cascade size distribution for  forest-fire model \cite{Drossel_1993} and number theoretic division model \cite{Luque_2008}, duration time distribution for critical branching process \cite{Zapperi_1995},  maximum velocity distribution of avalanches \cite{Goldenfeld_2012}, and size distribution for fragmentation of a discrete dimension rectangle into sticks (rectangles with unit width)\cite{Krapivsky}, to name a few.

The continued fraction is one of the way for expressing real numbers. Although the continued fraction has  extensively been studied in mathematics with several applications such as solution of quadratic equation and Pell's equation, it is also a topic of interest in the studies of dynamical systems \cite{CF_Dough_Hensely, Bosma}. 
In this paper, we show  the existence of scaling feature associated with continued fraction of a specific class of real numbers. The remarkable scaling can be viewed as a signature of Zipf's law, and the comprehensive numerical results adequately support this claim. Our arguments plausibly offer insight that the possible dynamical origin of such scaling could be SSR.

The scaling behavior in the present context can be easily recognized: Consider $X \in (0, 1)$ to be a random variable with uniform distribution. Let the simple continued fraction of the random number be denoted as $X = [0; k_1, k_2, \dots k_n]$. The limiting probability $P(k_n = k)$ is given by the so-called Gauss-Kuzmin distribution \cite{Gauss_Kuzmin}
\begin{equation}
\lim_{n \to \infty} P(k_n = k) = -\log_2\left(1-\frac{1}{(1+k)^2}\right).
\label{pdf_gk}
\end{equation}
The Eq.~(\ref{pdf_gk}) is also valid for a generic real number belonging to a set $\mathcal{R}$ (see Sec. II B). In fact, the asymptotic form of the distribution reduces to Zipf's law, $P(k) \sim k^{-2}$ for large $k$.  Despite this simplicity, the emergence of scaling behavior has not been adequately emphasized.  However, the numerical results clearly show the existence of the scaling, computed for sufficiently large $n \sim 10^6$ that offers a reasonable statistics, especially for large $k$.

The organisation of the paper is as follows: In Sec. II we recall the definition of continued fraction, algorithm, and interesting properties. A subsection is included here to mention relevant real numbers. The numerical results for probability distribution and spectral properties are presented in Sec. III. Section IV presents arguments that explain the emergence of scaling behavior. Finally, the paper concludes with a summary in Sec. V.

\section{Continued fraction}
A real number can be expressed in various ways. The most common way is {\it decimal representation}. In this case, the real number $X \in (0, 1)$ looks as
\begin{equation}
X = 0.d_1d_2\dots d_N,
\label{eq_dec}
\end{equation} 
where $d_i$ is the ith digit in the decimal expression. If the real number is irrational, then the corresponding rational approximation with $N$ precision is expressed by Eq.~(\ref{eq_dec}).

Besides the decimal representation, a mathematically more elegant expression is {\it continued fraction expansion}: 
\begin{equation}
X = x_0 + \frac{e_1}{x_1+\frac{e_2}{x_2+\dots}},
\end{equation}
where the elements $x_i$ and $e_i$ are positive integers. When $e_i = 1$, the general continued fraction reduces to simple continued fraction (SCF). In the remaining paper, the focus is only on the SCF. 
Due to typographical reason, the SCF is conveniently denoted as
\begin{equation}
X = [x_0; x_1, x_2, \dots, x_n],
\end{equation}
where $n$ is the number of elements in the expansion. Truncating the SCF expansion at $n$th step results in a convergent, which  provides a rational approximation of the real number.

The SCF is slightly efficient with respect to the decimal representation. This is a consequence of Lochs' theorem \cite{Lochs}:   For almost all $X$, $n< N$, as
\begin{equation}
\lim_{N\to \infty}\frac{n}{N} = \frac{6\ln2\ln10}{\pi^2} = 0.9702\dots.
\end{equation}

When an irrational number is terminated at finite precision, this may lead to an error in the computation of the SCF.  Express $\pi$ in the decimal representation $\pi = d_0.d_1d_2\dots d_N \dots = X+\delta$ such that $X<\pi$.
 Take $Y$ such that $X<\pi<Y$ and $Y-X = 10^{-N}$. The common terms in the SCF for both $X$ and $Y$ are in fact the elements of the  SCF for $\pi$.

Let us look at possible classes of the SCF expansion with striking example(s): 
\begin{itemize} 
\item Class-I (Finite): $18/7 = [2; 1, 1, 3]$. For rational numbers, the SCF expansion is finite. 
\item Class-II (Infinite with periodic or predictable pattern): $\sqrt{2} = [1; 2, 2, \dots]$, golden ratio: $\phi = [1; 1, 1, \dots]$, and $e = [2; 1, 2, 1, 1, 4, 1, 1, 6, 1, 1, 8, 1, 1, \dots ]$. For irrational numbers, the SCF expansion is nonterminating. In some cases, the pattern is either periodic or predictable. For $e$, the pattern repeats with a period of 3, except that 2 is added to one of the terms in each cycle.
\item Class-III (Infinite with random pattern): $\pi = [3; 7, 15, 1, 292, 1, 1, 1, 2, 1, 3, 1, \dots]$. Here, the nonterminating pattern appears irregular or random.
\end{itemize}

\subsection{Algorithms for computing the SCF}
For computing the SCF expansion of a real number, the commonly employed algorithms  are given below: 

(A1) Step-1: Start with $X_0 = X$. Step-2: Define $1/X_1 = X_0-\lfloor{X_0}\rfloor$, where $\lfloor \cdot \rfloor$ is the floor function. Step-3: Replace $X_0 \to X_1$ and go to Step-2. Iterate the process recursively until $\frac{1}{X_{n+1}} = 0$. Mathematically, the algorithm can be expressed as 
\begin{eqnarray}
X_0 = x_0 +\frac{1}{X_1}, {\rm where}~~ x_0 = \lfloor{X_0}\rfloor\nonumber\\
X_1 = x_1 + \frac{1}{X_2}, {\rm where}~~ x_1 = \lfloor{X_1}\rfloor\nonumber\\
\vdots \nonumber\\
X_n = x_n +\frac{1}{X_{n+1}}, {\rm where}~~ x_n = \lfloor{X_n}\rfloor.
\end{eqnarray}
The process terminates when $\frac{1}{X_{n+1}} = 0$.

(A2)  If the real number is  expressed as $X=p_0/q_0$, the Euclidean algorithm can be applied to compute the SCF:
\begin{eqnarray}
p_0 = x_0\cdot q_0 +r_0  \nonumber\\
(p_1 = q_0) = x_1 \cdot (q_1=r_0) + r_1 \nonumber\\
\vdots \nonumber\\
p_n = x_n\cdot q_n +r_n,
\label{eucl}
\end{eqnarray} 
where the quotient $x_i$'s are the elements of the SCF  and $r_i$'s are remainder. 
The process stops if $r_n = 0$. Interestingly $p_nq_{n-1}-p_{n-1}q_n = (-1)^n$.

(A3) The algorithm (A2) can be alternatively interpreted geometrically:  Start with a rectangle of dimension $\mathcal{A}_0 = p_0 \times q_0$. Construct $x_0$ squares of size $q_0$ and remove them. The remaining region forms a rectangle of area $\mathcal{A}_1 = p_1\times q_1 = q_0 \times r_0$. Again form $x_1$ squares of size $q_1 = r_0$ and cut them. The remaining area becomes a  rectangle of dimension $\mathcal{A}_2 = p_2\times q_2 = q_1 \times r_1$. The procedure is repeated until no rectangle  is left. Note that $\mathcal{A}_0 > \mathcal{A}_1 > \mathcal{A}_2 \dots > \mathcal{A}_n$.

\begin{figure}[b]
  \centering
  \scalebox{0.66}{\includegraphics{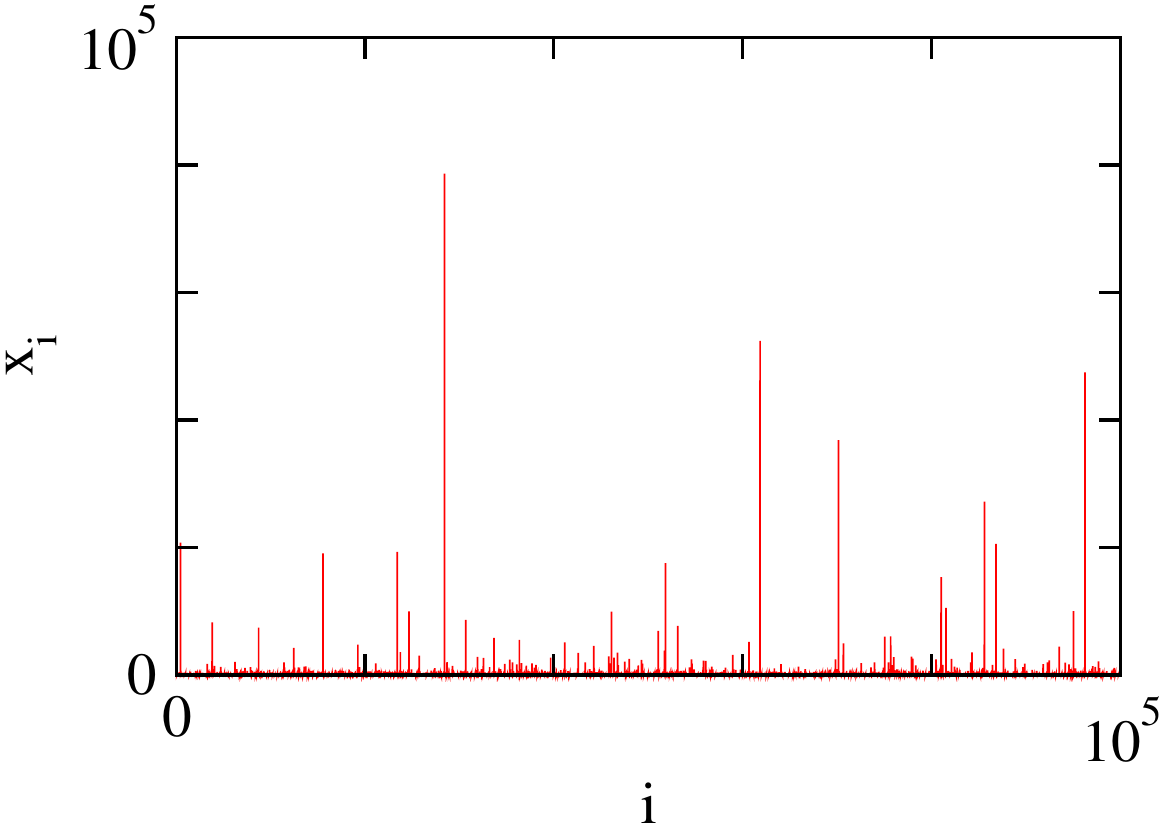}}
  \scalebox{0.66}{\includegraphics{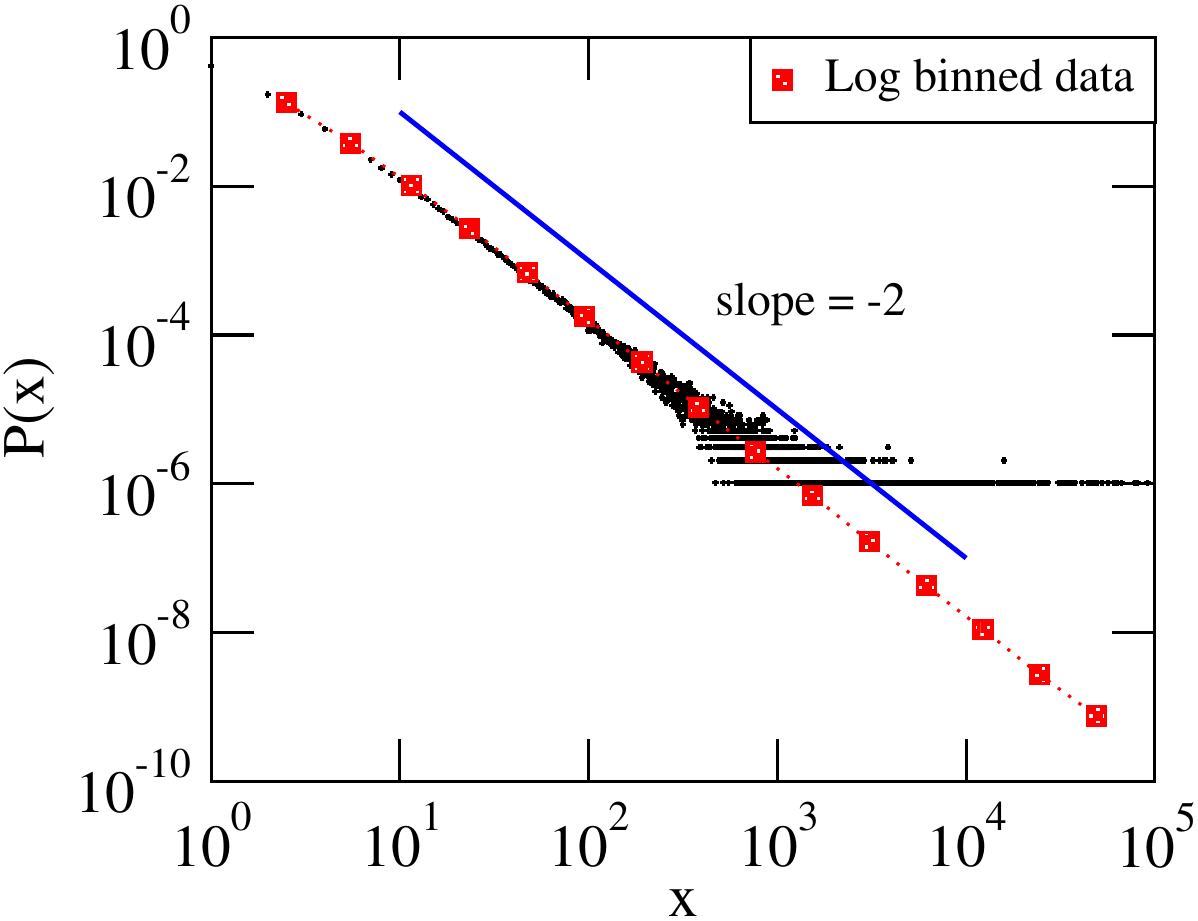}}
  \caption{Top panel: The plot of the SCF elements as a function of index for $\pi$ evaluated with precision $N=10^6$. Bottom panel: The probability distribution function $P(x)$. Filled square symbols show log binned data. Solid straight line with slope 2 is drawn for a comparison.}
 \label{fig01} 
\end{figure}

\begin{figure}[t]
  \centering
  \scalebox{0.66}{\includegraphics{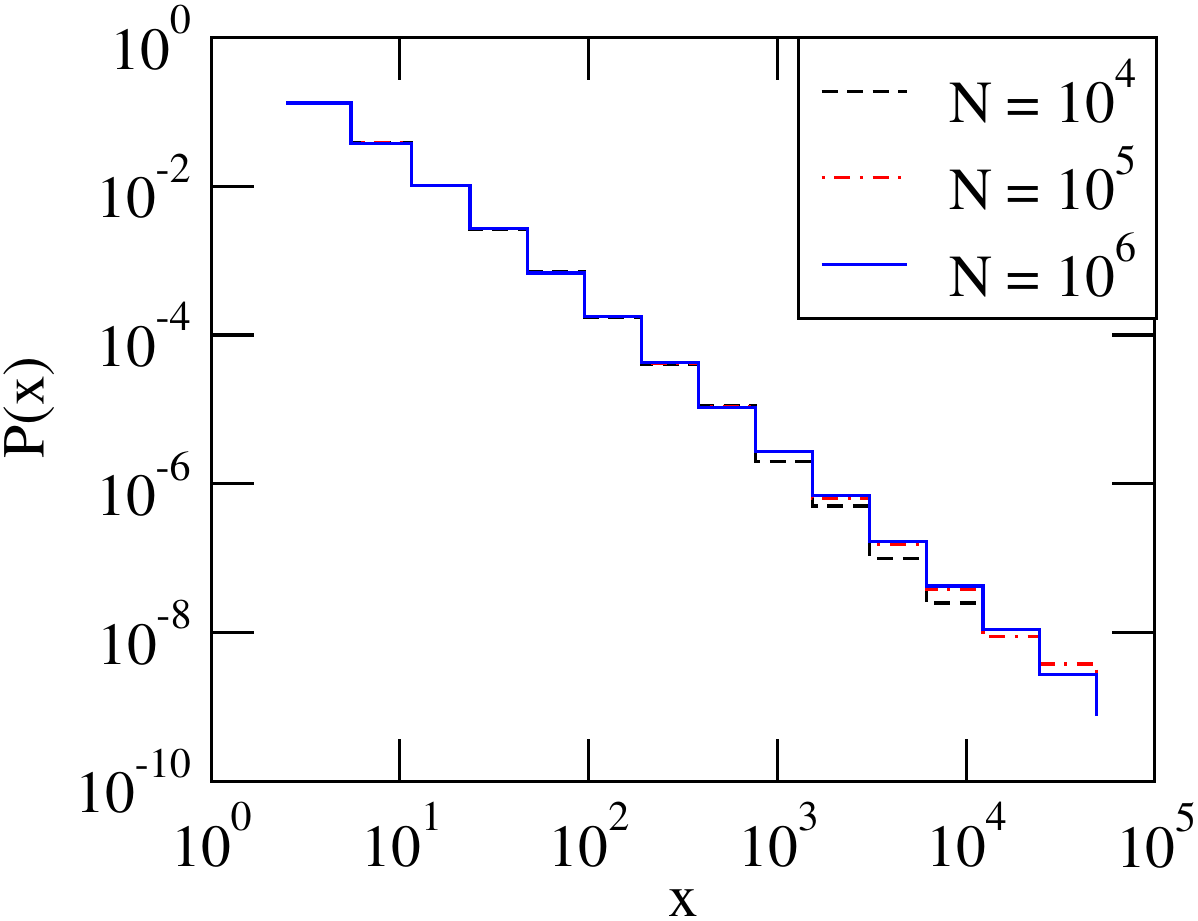}}
  \caption{Log binned PDF as histogram for the case of $\pi$ with different precision.}
 \label{fig02} 
\end{figure}

\begin{figure}[t]
  \centering
  \scalebox{0.66}{\includegraphics{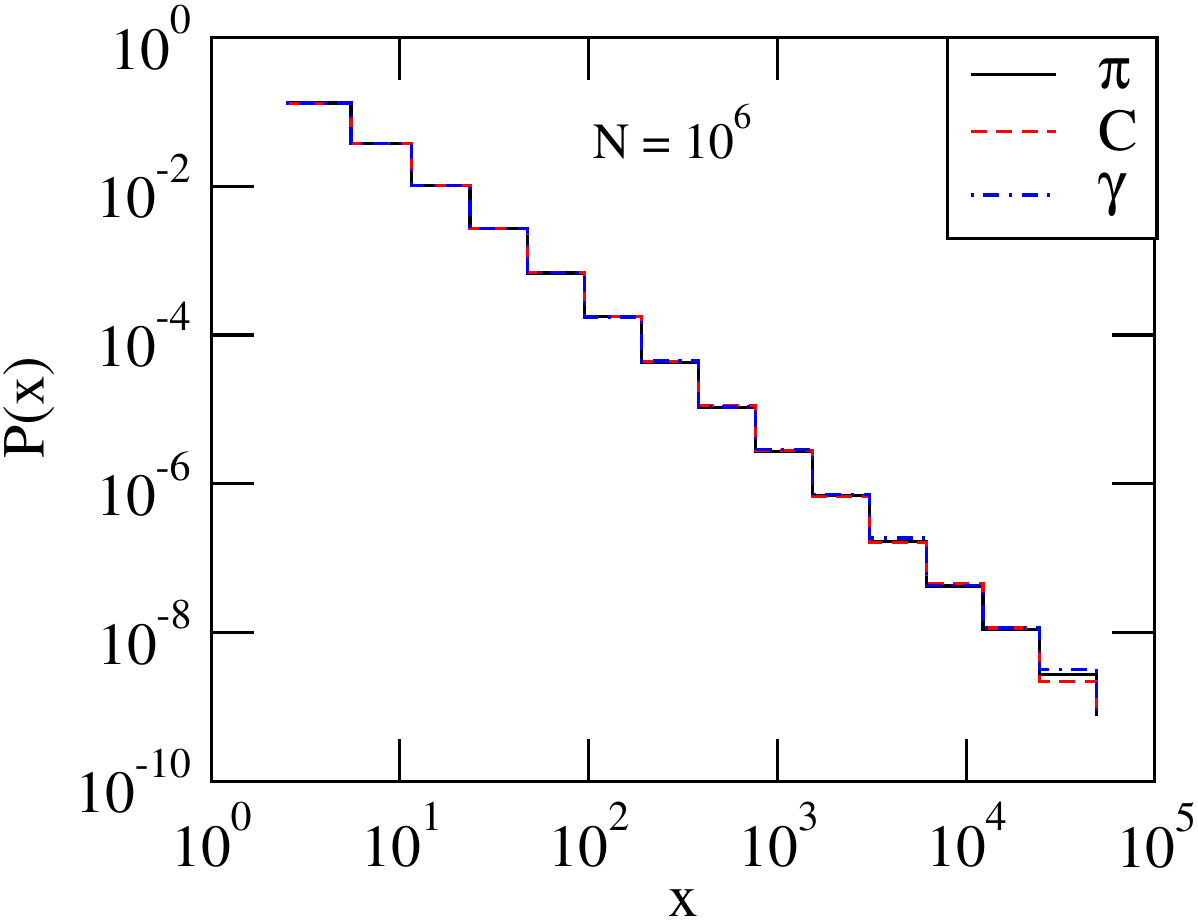}}
  \caption{The PDF for different real numbers $\pi$, Catalan, and Euler-Gamma.}
 \label{fig03} 
\end{figure}

\subsection{Examples of real numbers that belong to $\mathcal{R}$}
As shown in Sec. III, the characteristics of the SCF exhibit scaling feature for certain mathematical constants. It is useful to briefly mention a few. However, it is mathematically not known if the below mentioned constants are irrational.

\begin{itemize}

\item Catalan $C \approx 0.915966$:
\begin{equation}
C =  \sum_{k=0}^{\infty} (-1)^k  \frac{1}{(2k+1)^2}.
\end{equation}
 
\item Euler-Gamma $\gamma \approx 0.577216$:  
\begin{equation}
\gamma = \lim_{m\to \infty} \left( \sum_{k=1}^{m} \frac{1}{k}- \log m \right).
\end{equation}

\end{itemize}

In this paper, our main focus is on those real numbers for which the elements of the SCF appear to be random. Basically, these numbers belong to Class-III. Let us denote the set of such real numbers as $\mathcal{R}$.  It is noted that  the set includes certain irrational numbers and some mathematical constants: 
$\mathcal{R}\in \{\pi, a\pi+b, \pi^a, \log \pi,  \gamma, C, \sin(1) \dots\}$, where $a$ and $b$ are real numbers.

\section{Numerical Results}
Wolfram Mathematica software can be readily used to compute $\{x_i\}$, the elements of a real number in it's SCF representation. To compute the SCF, the following function is used in Mathematica: ContinuedFraction[$\cdot$], where  the real number with $m$ precision is computed with a function N[$\cdot, m]$. The methods of probability and statistics have been used to characterize the irregular pattern appearing in the sequence. 

First we focus on the probability distribution function (PDF) $P(x)$. Figure \ref{fig01} shows the plot of the elements with its index and the corresponding probability distribution. In order to estimate the critical exponent, more accurately, we present log binned data. In Fig.~\ref{fig02}, the same distribution is shown as histogram, where the precision varies from $N = 10^4$ to $10^6$. All different curves collapse onto a single curve, indicating robustness of the PDF  with respect to the precision. The same histogram is plotted in Fig.~\ref{fig03}, with different real numbers belonging to the set $\mathcal{R}$. This result reveals an interesting feature that the power law is a universal behavior. Moreover, the detailed numerical results, not shown, confirm emergence  of the scaling feature for irrational numbers constructed with $\pi$ by applying both linear [shift and scaling: $a\pi + b$] and nonlinear operations like \{$\pi^a, \log \pi$\}.

\begin{figure}[t]
  \centering
  \scalebox{0.66}{\includegraphics{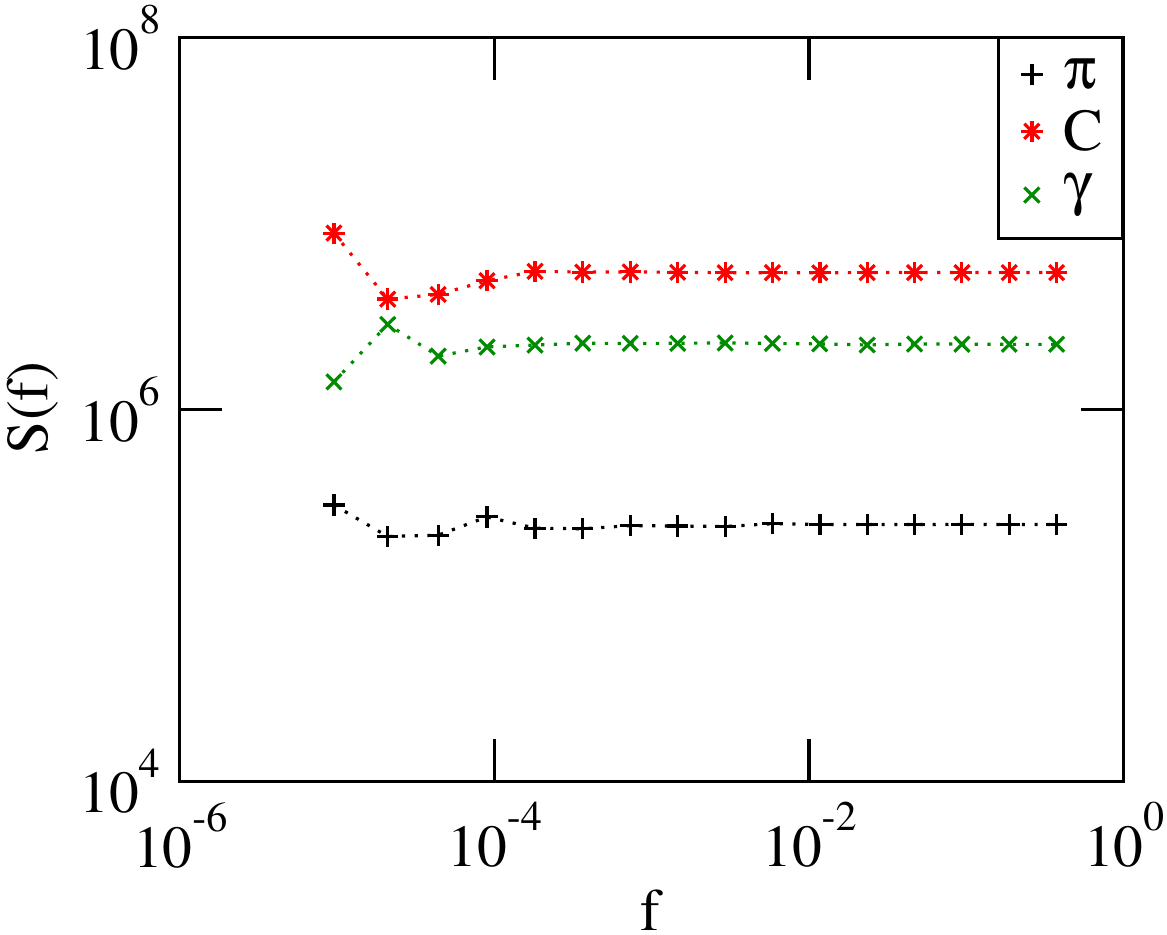}}
  \caption{The log binned power spectrum plot for $\pi, C,$ and $\gamma$.  The time series length is $2^{18}$.}
 \label{fig04} 
\end{figure}

\begin{figure}[t]
  \centering
  \scalebox{0.66}{\includegraphics{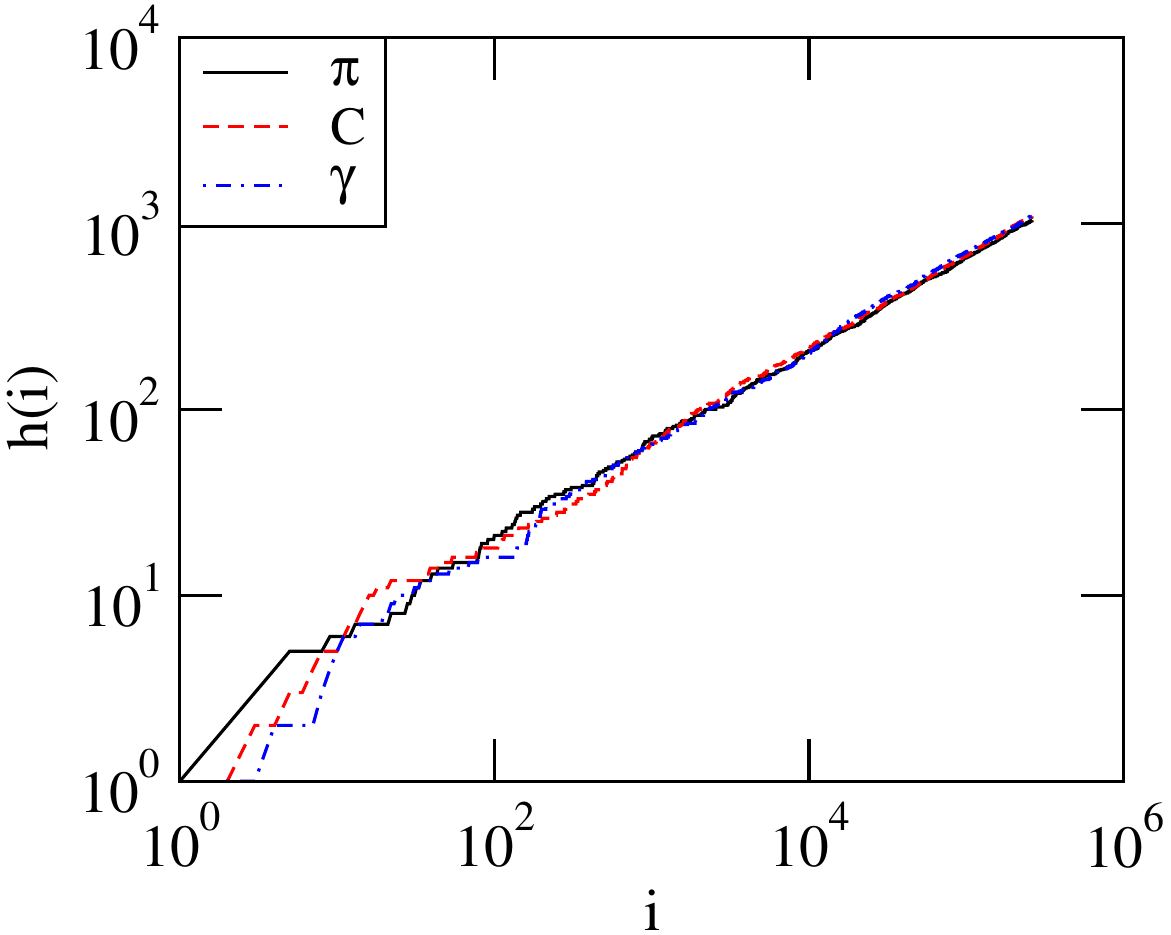}}
  \caption{The plot of $h(i)$ for different real numbers. The estimated value of the exponent is $\alpha \approx 0.56$.}
 \label{fig05} 
\end{figure}

We also compute power spectrum implementing Fast Fourier Transform method for the sequence of the elements, obtained from the SCF of the real numbers $\in \mathcal{R}$. Flat curves shown in Fig.~\ref{fig04} imply the elements are uncorrelated. Moreover, introduce $h(i)$  as the number of distinct elements appearing in the SCF.  If the entry $x_i$ occurring at step $i$ is distinct, then set the counter on as $\sigma_i = 1$. Thus, $h(i) = \sum_{j=0}^i \sigma_j$.  Figure \ref{fig05} shows the plot of $h(i)$ that grows sublinearly: $h(i)\sim i^{\alpha}$, with $\alpha$ being growth exponent.  The algebraic growth with non-trivial exponent is a signature of underlying scale-free distribution, while it may have a different form for other cases; for example, this eventually saturates for Bernoulli distribution.

\section{ Arguments for the scaling}
In order to understand the Zipf's law, heuristically, consider algorithm A1. Let $y_i = 1/X_i = X_{i-1}-\lfloor X_{i-1}\rfloor$. Note that $y_i \in (0, 1)$ and $X_i \approx x_i$. Clearly $x_i \sim 1/y_i$. Assuming $y_i$ to be a random variable with uniform distribution for $X\in \mathcal{R}$, and using probability chain rule, one can immediately see that $P(x) \sim x^{-2}$.

A correspondence can be invoked between the sample space reducing stochastic process and the simple continued fraction of a real number. The SSR can be easily understood in terms of fragmentation type events: Consider a stick of size $N$, and break it into two integer parts randomly and uniformly. Take one part, and repeat the breaking event successively. The process is iterated till the size becomes 1. The mapping is evident if the geometrical interpretation of the SCF (see algorithm A3) is viewed. Clearly, splitting a discrete rectangle into a smaller one successively, and eventually stopping when no rectangle can be formed, qualitatively describes the SSR mechanism. This mapping is useful in the sense that it can offer an explanation for the existence of Zipf's law associated with the SCF.

Writing Eq.~(\ref{eucl}) for subscript $i$ and multiplying by $q_i$, the size evolution of rectangle can be mathematically expressed as
\begin{equation}
\mathcal{A}_i = x_i \cdot \mathcal{S}_i + \mathcal{A}_{i+1},~~ {\rm with}~~ \mathcal{A}_{n} = 0 ~~ {\rm and}~~\mathcal{S}_n = 1, 
\end{equation}
where $\mathcal{S}_i$ is the area of square. Note that $\mathcal{A}_{i+1}<\mathcal{A}_{i}, \mathcal{S}_{i+1}< \mathcal{S}_{i}$, and $\mathcal{S}_{i}/\mathcal{A}_{i} \in (0, 1)$. The quantity in which we are interested is $x_i$:
\begin{equation}
x_i  =  \frac{\mathcal{A}_i - \mathcal{A}_{i+1}}{ \mathcal{S}_i} = \left \lfloor \frac{\mathcal{A}_i}{\mathcal{S}_i}\right\rfloor \approx \frac{1}{\mathcal{S}_i/\mathcal{A}_i}. 
\end{equation} 
Since $x_i$ is always positive, $\mathcal{S}_{i}>\mathcal{A}_{i+1}$. If $X\in \mathcal{R}$, it is assumed that $\mathcal{S}_{i}/\mathcal{A}_{i} $ behaves as uniformly distributed random variable. Consequently, the $x$ satisfies the power-law distribution. 

\section{ Summary}

In summary, we have revealed the existence of scaling or Zipf's law associated with the simple continued fraction of a class of real numbers that include a subset of irrational numbers and some mathematical constants. Our simple arguments offer an insight for the underlying mechanism, and it has been realized that the sample space reducing process may be responsible for the emergence of the scaling. Note that the subset of irrational numbers, for which the SCF exhibits randomness,  is non-denumerable. However, a remark can be drawn regarding the universality class: The observation of power-law with critical exponent 2 represents a universal  behavior.

\section*{ACKNOWLEDGMENT}
ACY  would like to acknowledge support through a grant ECR/2017/001702 funded by SERB, DST, Government of India.


\begin{thebibliography}{99} 
\bibitem{Newman_2005} M. E. J. Newman, Power laws, Pareto distributions and Zipf's law, Contemp. Phys. {\bf 46}, 323 (2005).

\bibitem{Sornette_2006} D. Sornette, {\it Critical Phenomena in Natural Sciences: Chaos, Fractals, Self-organization and Disorder: Concepts and Tools} (Springer, Berlin, 2006).

\bibitem{Miguel_2018} M. A. Mu\~noz, Colloquium: Criticality and dynamical scaling in living systems, \rmp{90}{031001}{2018}.




\bibitem{Bak_1996} P. Bak, {\it How Nature Works: The Science of Self Organized Criticality} (Copernicus Press, New York, 1996).

\bibitem{Bak_1987} P. Bak, C. Tang, and K. Wiesenfeld, Self-organized criticality: An explanation of the $1/f$ noise, \prl{59}{381}{1987}.
\bibitem{Yadav_2012} A. C. Yadav, R. Ramaswamy, and D. Dhar, Power spectrum of mass and activity fluctuations in a sandpile, \pre{85}{061114}{2012}.



\bibitem{Levina_2007} A. Levina, J. M. Herrmann, and T. Geisel, Dynamical synapses causing self-organized criticality in neural networks, \natphys{3}{857}{2007}.

\bibitem{Manchanda_2013} K. Manchanda,  A. C. Yadav, and R. Ramaswamy, Scaling behavior in probabilistic neuronal cellular automata, \pre{87}{012704}{2013}.

\bibitem{Jonathan_2017} J. Touboul and A. Destexhe, Power-law statistics and universal scaling in the absence of criticality, \pre{95}{012413}{2017}.



\bibitem{Djordje_2011} D. Spasojevi{\'c}, S. Jani{\'c}evi{\'c}, and M. Kne{\~z}evi{\'c}, Numerical evidence for critical behavior of the two-dimensional nonequilibrium Zero-Temperature Random Field Ising Model, \prl{106}{175701}{2011}.

\bibitem{Knecht_2012} C. L. Knecht, W. Trump, D. ben-Avraham, and R. M. Ziff, Retention Capacity of Random Surfaces, \prl{108}{045703}{2012}.

\bibitem{Munn_2018} B. Munn and P. Gong, Critical dynamics of natural time varying images, \prl{121}{058101}{2018}.

\bibitem{Kinouchi_2006} O. Kinouchi and M. Copelli, Optimal dynamical range of excitable networks at criticality, \natphys{2}{348}{2006}.



\bibitem{Redner_2001} S. Redner, {\it A Guide to First Passage Processes} (Cambridge University Press, Cambridge, 2001).

\bibitem{Dickman_2001} R. Dickman and D. ben-Avraham, Continuously variable survival exponent for random walks with movable partial reflectors, Phys. Rev. E 64, 020102(R) (2001).
\bibitem{Avinash_2018} A. C. Yadav, Critical P{\'o}lya urn, \pre{98}{022119}{2018}.



\bibitem{Albert_1999} A.-L. Barab{\'a}si and R. Albert, Emergence of scaling in random networks, \sc{286}{509}{1999}.


\bibitem{Levy_1996} M. Levy and S. Solomon, Spontaneous scaling emergence in generic stochastic system, Int. J. Mod. Phys. C {\bf 7}, 745 (1996).

\bibitem{Takayasu_1997} H. Takayasu, A.-H. Sato, and M. Takayasu, Stable infinite variance fluctuations in randomly amplified Langevin systems, \prl{79}{966}{1997}.

\bibitem{Yamamoto_2012} K. Yamamoto and Y. Yamazaki, Power law behavior in a cascade process with stopping events: A solvable model, \pre{85}{011145}{2012}.

\bibitem{Yamamoto_2014} K. Yamamoto, Stochastic model of Zipf's law and the universality of the power-law exponent, \pre{89}{042115}{2014}.




\bibitem{Murtra_2015} B. Corominas-Murtra, R. Hanel, and S. Thurner, Understanding scaling through history-dependent processes with collapsing sample space, \pnas{112}{5348}{2015}.

\bibitem{Murtra_2017} B. Corominas-Murtra, R. Hanel, and S. Thurner, Sample space reducing cascading
processes produce the full spectrum of scaling exponents, Scientific Reports {\bf 7}, 11223 (2017).

\bibitem{Murtra_2018} B. Corominas-Murtra, R. Hanel, L. Zavojanni, and S. Thurner, How driving rates determine the statistics of driven non-equilibrium systems with stationary distributions, Scientific Reports {\bf 8}, 10837 (2018).

\bibitem{Yadav_2016} A. C. Yadav, Survival-time statistics for sample space reducing stochastic processes, \pre{93}{042131}{2016}.

\bibitem{Yadav_2017} A. C. Yadav,  Correspondence between a noisy sample space reducing process and records
in correlated random events, \pre{96}{032134}{2017}.






\bibitem{Drossel_1993} B. Drossel, S. Clar, and F. Schwabl, Exact results for the one-dimensional self-organized critical forest-fire model, \prl{71}{3739}{1993}.


\bibitem{Luque_2008} B. Luque, O. Miramontes, and L Lacasa, Number theoretic example of scale-free topology inducing self-organized criticality, \prl{101}{158702}{2008}.

\bibitem{Zapperi_1995} S. Zapperi, K. B. Lauritsen, and H. E. Stanley, Self-organized branching processes: Mean-field theory for avalanches, \prl{75}{4071}{1995}.


\bibitem{Goldenfeld_2012} M LeBlanc, L. Angheluta, K. A. Dahmen, and N. Goldenfeld, Distribution of maximum velocities in avalanches near the depinning transition, \prl{109}{105702}{2012}.

\bibitem{Krapivsky} E. Ben-Naim and P. L. Krapivsky, Jamming and tiling in fragmentation of rectangles, arXiv:1905.06984.




\bibitem{CF_Dough_Hensely} D. Hensley, {\it Continued Fractions} (World Scientific, 2006).  

\bibitem{Bosma} W. Bosma et. el., {\it Continued Fractions} (Lecture Notes, 2012-13), \\ https://www.math.ru.nl/~bosma/Students/CF.pdf.

\bibitem{Gauss_Kuzmin}  E. W. Weisstein,  Gauss--Kuzmin Distribution, MathWorld, \\ http://mathworld.wolfram.com/GaussKuzminDistribution.html.

\bibitem{Lochs}  E. W. Weisstein,  Lochs' Theorem, MathWorld, http://mathworld.wolfram.com/LochsTheorem.html.


\end{thebibliography}
\end{document}